\documentclass[a4paper,fleqn,usenatbib]{mnras}

\usepackage{newtxtext,newtxmath}
\usepackage[T1]{fontenc}
\usepackage{ae,aecompl}

\usepackage{graphicx}	% Including figure files
\usepackage{amsmath}	% Advanced maths commands
\usepackage{amssymb}	% Extra maths symbols
\usepackage{subfigure}
\usepackage{multirow}
\usepackage{ulem}
\usepackage{url}
\usepackage{natbib}
\usepackage{caption}
\usepackage{lipsum}
\title[AstroCatR]{AstroCatR: a Mechanism and Tool for Efficient Time Series Reconstruction of Large-Scale Astronomical Catalogues}

\author[Ce Yu et al.]{
	Ce Yu,$^{1}$ $^{2}$\thanks{E-mail: yuce@tju.edu.cn (CY)}
	Kun Li,$^{1}$ $^{2}$
	Shanjiang Tang,$^{1}$ $^{2}$
	Chao Sun,$^{1}$ $^{2}$
	Bin Ma,$^{3}$
	and Qing Zhao$^{4}$
	\\
	$^{1}$ College of Intelligence and Computing, Tianjin University, No.135 Yaguan Road, Haihe Education Park, Tianjin 300350, China\\
	$^{2}$ NAOC-TJU Joint Research Center in Astro-Informatic, No.135 Yaguan Road, Haihe Education Park, Tianjin 300350, China\\
	$^{3}$ National Astronomical Observatories, Chinese Academy of Sciences, No.20 Datun Road, Chaoyang District, Beijing 100012, China\\
	$^{4}$ School of Computer Science and Information Engineering, Tianjin University of Science $\&$ Technology, Tianjin 300457, China
}

\date{Accepted 2020 May 18. Received 2020 May 17; in original form 2019 September 23}

\pubyear{2020}

\begin{document}
	
	\label{firstpage}
	\pagerange{\pageref{firstpage}--\pageref{lastpage}}
	\maketitle
	
	\begin{abstract}
		Time series data of celestial objects are commonly used to study valuable and unexpected objects such as extrasolar planets and supernova in time domain astronomy. Due to the rapid growth of data volume, traditional manual methods are becoming extremely hard and infeasible for continuously analyzing accumulated observation data. To meet such demands, we designed and implemented a special tool named AstroCatR that can efficiently and flexibly reconstruct time series data from large-scale astronomical catalogues. AstroCatR can load original catalogue data from Flexible Image Transport System (FITS) files or databases, match each item to determine which object it belongs to, and finally produce time series datasets. To support the high-performance parallel processing of large-scale datasets, AstroCatR uses the extract-transform-load (ETL) preprocessing module to create sky zone files and balance the workload. The matching module uses the overlapped indexing method and an in-memory reference table to improve accuracy and performance. The output of AstroCatR can be stored in CSV files or be transformed other into formats as needed. Simultaneously, the module-based software architecture ensures the flexibility and scalability of AstroCatR. We evaluated AstroCatR with actual observation data from The three Antarctic Survey Telescopes (AST3). The experiments demonstrate that AstroCatR can efficiently and flexibly reconstruct all time series data by setting relevant parameters and configuration files. Furthermore, the tool is approximately 3X faster than methods using relational database management systems at matching massive catalogues.
	\end{abstract}
	
	
	%%%%%%%%%%%%%%%%%%%%%%%%%%%%%%%%%%%%%%%%%%%%%%%%%%
	
	%%%%%%%%%%%%%%%%% BODY OF PAPER %%%%%%%%%%%%%%%%%%

	\section{Introduction}
	%
	%This is a simple template for authors to write new MNRAS papers.
	%See \texttt{mnras\_sample.tex} for a more complex example, and \texttt{mnras\_guide.tex}
	%for a full user guide.
	%
	%All papers should start with an Introduction section, which sets the work
	%in context, cites relevant earlier studies in the field by \citet{Others2013},
	%and describes the problem the authors aim to solve \citep[e.g.][]{Author2012}.

	\label{sec_1_introduction}
	
	Time series data extracted from catalogues are essential for the analysis of the period and characteristics of celestial objects in time domain astronomy (TDA)\footnote{\textsl{This is a pre-copyedited, author-produced PDF of an article accepted for publication in} Monthly Notices of the Royal Astronomical Society \textsl{following peer review. The version of record} Ce Yu, Kun Li, Shanjiang Tang, Chao Sun, Bin Ma, and Qing Zhao, AstroCatR: a Mechanism and Tool for Efficient Time Series Reconstruction of Large-Scale Astronomical Catalogues, Monthly Notices of the Royal Astronomical Society, May 2020, \textsl{is available online at} \url{https://doi.org/10.1093/mnras/staa1413}}. Recent advances in observation technology and the increasing number of astronomical observation facilities are providing extremely rich data resources for such time-domain astronomy research. For instance, Gaia DR2 released astronomical parameters for 160 million sources and more than 500,000 variable stars~\citep{marrese2019gaia}. NASA$'$s Transiting Exoplanet Survey Satellite (TESS)~\citep{Ricker2015Transiting} mission is an all-sky survey that will discover thousands of exoplanets around nearby bright celestial objects. The Large Synoptic Survey Telescope (LSST) will produce raw imaging data at a rate of 15 TB/night and will collect over 50 PB for the catalogue data~\citep{ivezic2019lsst}. More advanced telescopes are already under construction, including the Thirty Meter Telescope (TMT)~\citep{TMT}, European Extremely Large Telescope (E-ELT)~\citep{ELT}, Giant Magellan Telescope (GMT)~\citep{GMT}, and the James Webb Space Telescope (JWST)~\citep{JWST}.%The growth of astronomical data complicates the time series reconstruction of celestial objects, which is currently important for large-scale astronomical catalogues in the scientific and astronomical data processing fields.  
	
	For optical astronomical observations, catalogue data are generated from photometry processing on original images. Typically, the objects detected and measured in the same image are listed in a single catalogue file. This type of data organization is not suitable for TDA research because the time series information of each object is indirect. Although we can retrieve the time series data of the specific candidates via cone search or cross-matching, as shown in Figure 1, the efficiency will continue to decrease as the volume of data continues to increase. In addition, some valuable variable objects outside the existing man-made candidate lists will be missed. 
	%The Figure 1 demonstrates the main work of this paper. According to the information given by astronomical catalogues, the same celestial objects are marked by matching calculation, so as to obtain the time series data and scatter plots of all celestial objects.
	
	\begin{figure*}
		\centering
		\includegraphics[width=0.75\textwidth]{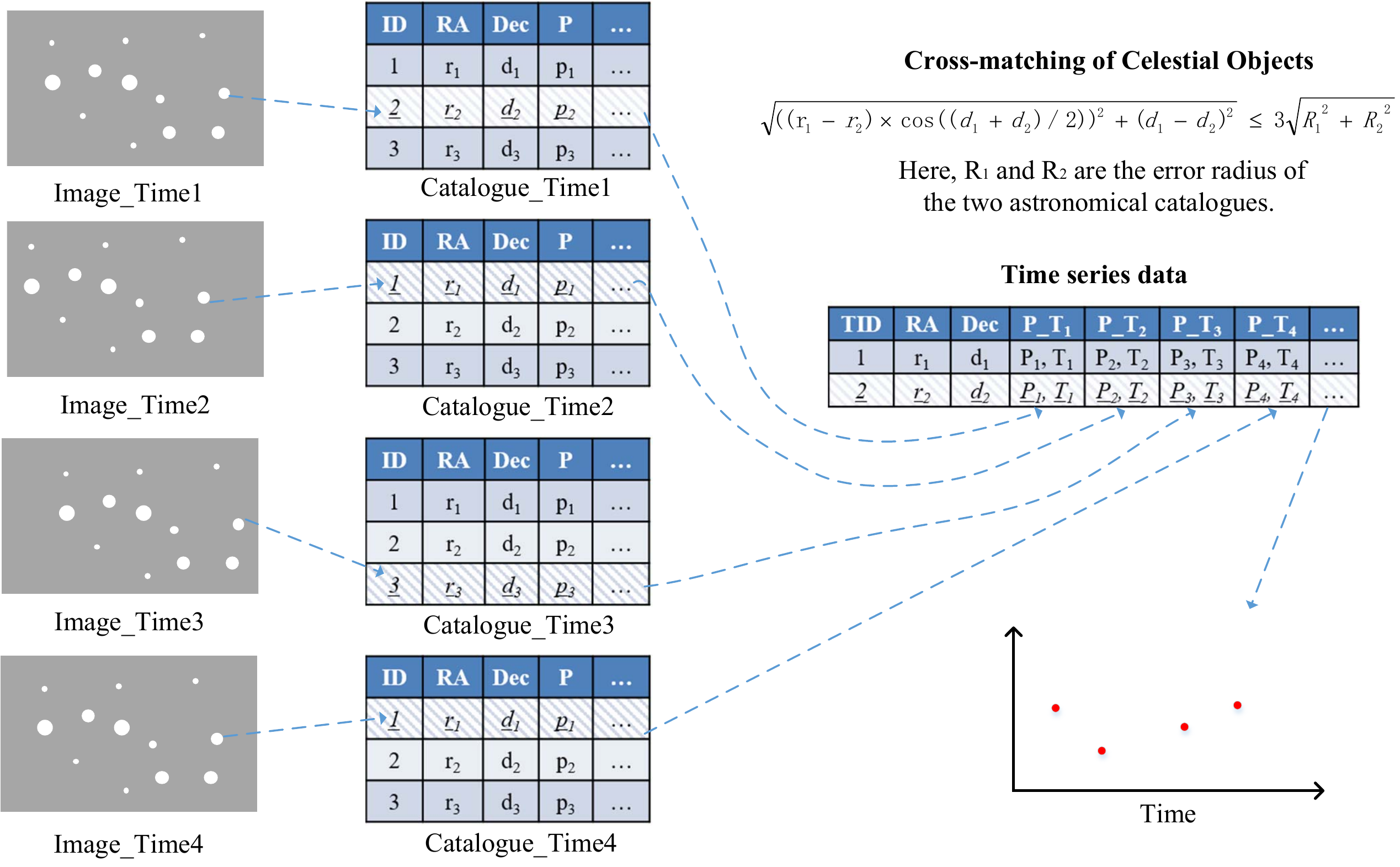}
		\caption{\small The time series reconstruction extracts meta data from astronomical catalogues and obtains the information of the same celestial objects through matching calculation. Time series data can be represented by scatter plots, which provides a basis for subsequent research on fitting and classifying light curves.}
		\label{fig_1}
	\end{figure*}
 
	Rapidly developing AI technologies provide an opportunity to extend the time series data analysis to the entire dataset, which promotes more exciting discoveries. Valuable and complex time-domain information can be obtained by automatically clustering or classifying time series data. For example, two exoplanets were discovered by the combination of Google AI and data from the Kepler space telescope~\citep{shallue2018identifying}. However, before being able to take full advantage of AI technologies, we have to prepare the appropriate form of the dataset.
	
	%Notebly, reconstructing time series data from astronomical catalogues is a significant pre-processing step in astronomical data analysis. Furthermore, important astrophysical topics, such as the discovery of supernovae and extrasolar planets, the evolution of star-forming regions, and the formation of galaxy, could be addressed by monitoring and studying time series at different wavelengths. In the near future, data-driven Artificial Intelligence (AI) will prompt the design of a learning mechanism and an automatic algorithm to model massive amounts of the time series data. Therefore, an effective mechanism is needed to reconstruct the time series data of all celestial objects from astronomical catalogues~\citep{AstroCloud}. 
 
	To solve the above problems, we designed and implemented a special tool named AstroCatR to construct time series data for each object in the entire dataset from the original catalogue files, each of which corresponds to a single observation image. The output of AstroCatR is named TSCat, which is a list that merges all the celestial objects, as well as their corresponding time series data, from the inputted catalogue files. Basically, AstroCatR needs to iterate each object in each inputted catalogue file and cross-match it with TSCat to determine the following operation. Performance and accuracy are the most considerable challenges for AstroCatR to process large-scale catalog datasets. 	
	
	The key features of AstroCatR include:
		
	$\bullet$ High performance parallel processing. The extract-transform-load preprocessing module is designed to create sky zone files and balance the workload.
	
	$\bullet$ TS-Matching algorithm. The matching module is based on the overlapped indexing method and an in-memory reference table to improve the accuracy, and uses multiprocess parallel technology to improve the performance.
	
	$\bullet$ Usability and scalability. AstroCatR is released as a ready-to-use open source tool, and its module-based software architecture ensures flexibility and scalability. The output can be stored in CSV files or transformed into other formats as needed.
	
	The remainder of this paper is organized as follows: Related works on matching calculation and storage of the astronomical catalogue are introduced in Section 2. Section 3 presents the architecture of AstroCatR and the details of the modules and algorithms. Section 4 discusses evaluation of AstroCatR with real catalogue data. In the final section, we summarize the work and discuss the future work.
	
	\section{Related Work}
	\label{sec_2_Related Work}
	
	The foundation for reconstructing time series data is the astronomical catalogue matching calculation to determine whether two records describe the same celestial object. The key to the matching calculation performance is the storage and access to catalog data.
	
	\subsection{Matching Calculation}
	\label{subsec_2.1_Matching Calculation}
	
	The matching calculation module of this paper is an improvement on the basis of cross-matching. Traditional cross-matching compares the data of two catalogues. The reference table is designed to assist the matching calculation of AstroCatR. Nevertheless, we can learn from cross-matching technology and optimization methods. The criterion for location-based cross-matching is the approximate coincidence of celestial coordinates~\citep{yu2019astronomical}. To increase the speed of cross-matching calculations, multiple technologies have been employed, including high-performing computing (HPC) and sky partitioning. There are acceleration methods based on MPI~\citep{Zhao2009A}, the multi GPU environment~\citep{Budavari2013Xmatch}, the CPU-GPU cluster~\citep{Jia2015Cross}, the Hadoop ecological system~\citep{li2014optimizing}, and Spark framework~\citep{zečević2019axs}. Additionally, various cross-matching tools have been recently developed to handle massive catalogues, such as ARCHES~\citep{Motch2016The}, C$^3$~\citep{Riccio2017C3}, and catsHTM~\citep{Soumagnac2018catsHTM}. The use of sky partitioning and parallel processing methods is accompanied by the emergence of boundary problems. The methods to address the problems include increasing redundant data on the boundary~\citep{Zhao2009A} and using two indexes~\citep{Peng2014New}. However, most of the aforementioned cross-matching works are less appropriate for direct application to time series data reconstruction. The matching process in this work is to obtain a full list of all objects (each object followed by corresponding time series data) for the catalogues from the same observation device during the specific period. To enable such process, we need a refence table to maintain the list and improve the computing performance. The detailed discussion of the matching method is in Subsection 3.3.

	\subsection{Storage and Access of Astronomical Catalogue}
	\label{subsec_2.2_Storage and Access of Astronomical Catalogue}
	
	Generally, the processing of astronomical catalog data is based on databases, but most of the catalogue data will not be accessed in matching calculation process; only the reference table has large-scale read and write operations. The key to improving the performance of time series data reconstruction lies in the storage and access of the reference table. Traditional database management systems (DBMS) and Hadoop ecosystems~\citep{Richter2014Towards} must load data before querying. Loading operations take up a considerable proportion of the overall execution time~\citep{idreos2011here} and they also create a large amount overhead space. With respect to the NoSQL method, original data must inevitably be imported into the dedicated database or file system, and this importation will require a significant overhead of space and time~\citep{hong2016aquadexim}. Most of these methods optimize the memory access of the DBMS querier, which restricts the choice and expansion of the computing architecture. \cite{alagiannis2012nodb} proposed an adaptive indexing mechanism and flexible cache structure for providing effective access to original data files. NoDB regards original data files as the best source of DBMS. AstroCatR adopts the NoDB strategy, but because of the intermediate output among the processing processes, the output data can also be processed by using databases as required.

	\section{Efficient Time Series Reconstruction System}
	\label{sec_3_Efficient Time Series Reconstruction System}

	\subsection{Software Architecture of AstroCatR}
	\label{subsec_3.1_Software Archicture of AstroCatR}
	
	AstroCatR aims to efficiently reconstruct time series data from massive astronomical catalogues, where the performance and accuracy are the most important issues. Considering the ease of use and scalability of the software system, AstroCatR is divided into three independent modules as shown in Figure 2.
	
	%This work focuses on efficiently reconstructing the time series data from massive homologous astronomical catalogues for all celestial objects. AstroCatR transforms original flexible image transport system (FITS) files into sky zoning files. \textbf{A partition function is designed to maintain workload balance. We leveraged an improved approach to accelerate and promote the accuracy of matching calculations using reference tables, magnitude information, and mixed indexes. The software archicture of AstroCatR is shown in Figure 2. Each component handles only a part of the whole system, effectively reducing the scale and complexity of the problem.} 
	
	\begin{figure}
		\centering
		\includegraphics[width=0.5\textwidth]{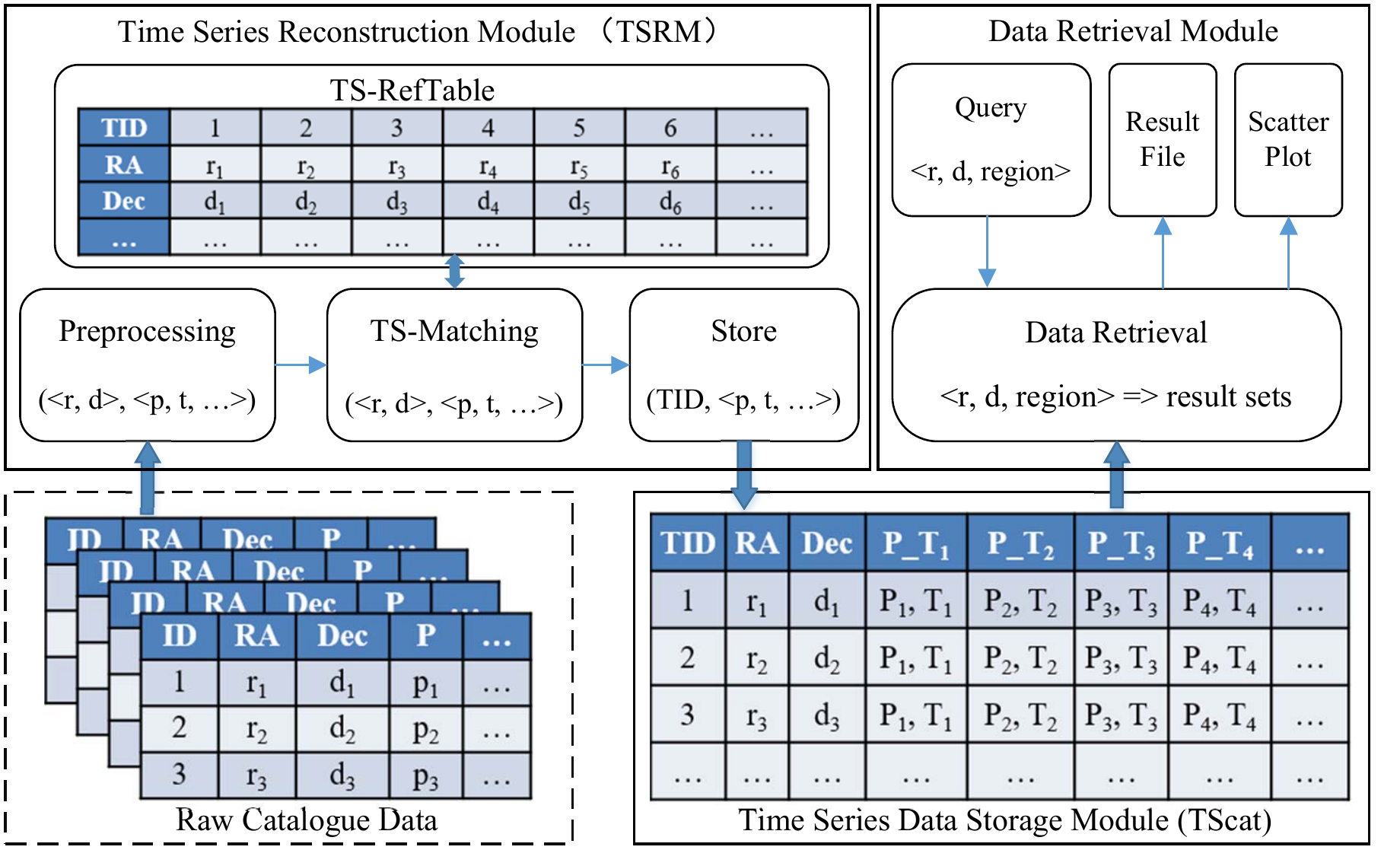}
		\caption{\small AstroCatR overview. The software architecture is divided into three components: reconstruction module, time series data storage, and data retrieval module.}
		\label{fig_2}
	\end{figure}
	
	$\bullet$ Time Series Data Storage (TSCat)
		
	TSCat is a specially designed data structure for storing time series data. The default format is CSV, which is imported into the database according to the needs of users for further research. Data of the same celestial object exist in the same file. Each row in TSCat represents the information of a celestial object, including position, index, magnitude, error of magnitude, and observation time. The incremental catalogues can be transformed into sky zoning files, and can match celestial objects using the existing reference table for the same sky zone. The incremental output can be saved to the corresponding TSCat file block.
	
	$\bullet$ Reconstruction Module (TSRM)
	
	TSRM is responsible for reading the original catalog data, determining the object to which each record belongs, and restoring the time series data into TSCat. As the most complex module of AstroCatR, TSRM is divided into three submodules: 1)  \textit{Preprocessing} submodule. The first is extract-transform-load (ETL) and partition processing of the original astronomical catalogue data. The preprocessing extracts the information needed to reconstruct the time series and divides the data into different sky zoning files. 2)  \textit{TS-Matching} submodule. The TS-Matching calculation algorithm is designed for homologous catalogues with in-memory reference tables. The boundary problems are solved by the overlapped indexing method instead of redundancy. 3) \textit{Store} submodule. The time series data are written to TSCat in the specified format.
	
	$\bullet$ Data Retrieval Module
	
	AstroCatR provides query services for the reconstructed time series data. The definition of the query is described as follows: given the position information of a query, such as right ascension (RA) and declination (DEC), find all time series data that represent the coordinates matched. The query request is submitted, and returns target data and corresponding light curves (scatter plots) from the time series datasets of the astronomical catalogues. 
	
	AstroCatR is a command-line opensource program running on the Linux platform, which is implemented in C and Python. Its capabilities are based on specialized sky partitioning and MPI parallel programming. It is designed to deal with massive catalogues with maximum user flexibility given to users in terms of parameter setting and catalogue formats. Table 1 presents the basic information and runtime requirements of AstroCatR, including environment requirements, deployment and installation information, usage methods, and data input and output formats. Each AstroCatR module is configured with separate parameters.
	
	\begin{table}
		\centering
		\caption{Basic Information on AstroCatR}
		\begin{tabular}{lp{3cm}}
			\hline\noalign{\smallskip}
			Basic features of AstroCatR                         & Notes                                                                                                                           \\ \hline
			\multirow{7}{*}{Operating Environment Requirements} & Linux Operating System                                                                                                          \\
			& C and C++ Compilers                                                                                                             \\
			& Python \textgreater{}= 2.7                                                                                                      \\
			& Git Client(1.8 or greater)                                                                                                      \\
			& MPI (Required for parallel processing) \\
			& Gnuplot                                                                                                                         \\
			& Cfitsio                                                                                                                         \\ \\
			Deployment and Installation                         & Makefile                                                                                                                        \\ \\
			\multirow{3}{*}{Usage Method}                       & Shell Scripts (All batch-related programs)                                                                                      \\
			& mpirun (MPI)                                                                                                                 \\
			& Python Script(Query)                                                                                                            \\ \\
			Input Data                                          & Catalogue FITS Files                                                                                                                      \\
			Output Data                                         & CSV Files          \\  \hline                                                                                                        
		\end{tabular}
	\end{table}

	The above algorithms and source codes of AstroCatR are published at \url{https://gitee.com/AstroTJU/AstroCatR}. The running environment of AstroCatR is the Linux operating system, and has the following dependencies: Mpich, Python, Gnuplot and Cfitsio. Cfitsio~\citep{pence1999cfitsio} is used to parse catalogue FITS file information, MPI is used to accelerate TS-Matching calculations and Python is used to manage user queries. The main third-party tools used by AstroCatR are hierarchical equal area isoLatitude pixelation (HEALPix)~\citep{HEALPix} and hierarchical triangular mesh (HTM)~\citep{HTM}, which perform fine-grained partitioning of celestial sphere surfaces.  
	
	\subsection{Parallel Processing Support for Large-Scale Datasets}
	\label{subsec_3.2_Parallel Processing Support for Large-Scale Datasets}
	
	The process of reconstruction is shown in Figure 3. Each record needs to be identified to know which object it belongs to, and it is similar to the traditional cross-matching between two catalogues. Pseudospherical indexes such as HEALPix can reduce the computational complexity, but cross-matching remains a hindrance influencing the consumption of time when processing large-scale astronomical catalogues~\citep{Riccio2017C3}. In the reconstruction of time series, the matching calculation takes place between frequently sampled multiple catalogues, so the computational quantity is larger. To support parallelization of the TS-Matching procedure, AstroCatR provides an extract-transform-load preprocessing module and partition function for implementing data partitioning and load balancing.
	
	\begin{figure}
		\centering
		\includegraphics[width=0.45\textwidth]{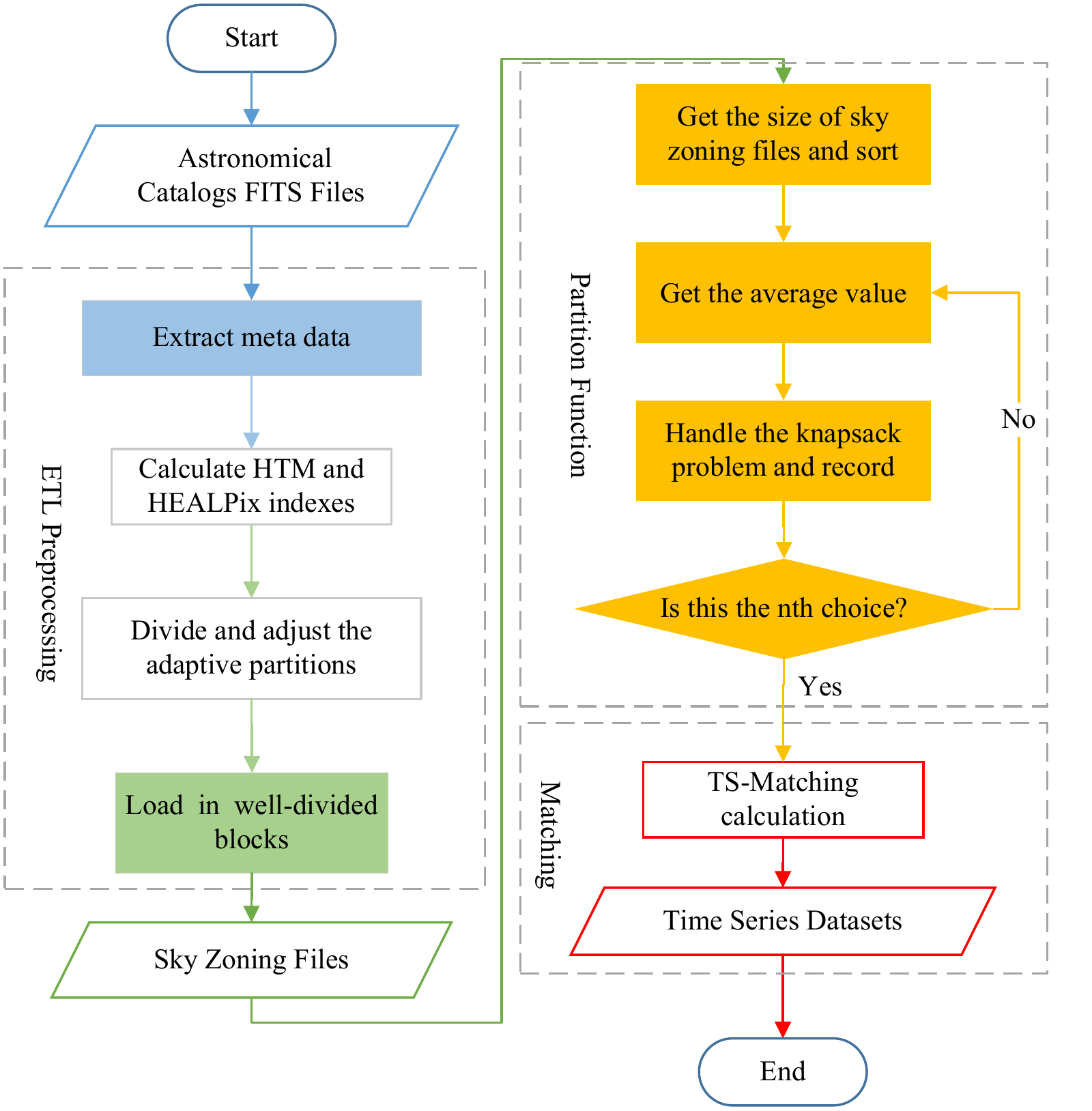}
		\caption{\small Data processing flow for time series reconstruction. 1) The ETL preprocessing program loads the original catalogues and extracts the meta data from catalogue FITS files and then generates sky zoning files. The data are grouped according to sky zones, which are divided by the spatial index to avoid invalid match calculations. 2) The partition function optimizeoptimizes the distribution of sky zoning files to balance the workload. 3) The TS-Matching calculation program marks celestial objects.}
		\label{fig_3}
	\end{figure}	
	
	$\bullet$ Preprocessing and Partitioning
	
	The extract-transform-load preprocessing is to extract and process information that is used to study time series data from the original astronomical catalogues. HTM and HEALPix indexes are calculated using the extracted ${RA}$ and ${DEC}$, and the partition level is adjusted according to the specific data distribution conditions. The indexes can reduce TS-Matching operation computations, and overlapping using the two indexes solves the boundary problems mentioned in Subsection 3.3. Metadata are extracted from catalogue FITS files, and Table 2 provides the names, attributes and explanations of the meta data.
	
	The data are then loaded into the divided sky blocks for further processing, and these data files are named sky zoning files. We use multiple processes to manage sky zoning files, so tasks must be divided, and the problem is converted into a step-by-step knapsack problem. The definition of the knapsack problem is given below. Tasks must be partitioned using parallel processing methods, and the most time-consuming task determines the final completion time. To limit the size and number of sky zoning files, we tune-up the partition by adjusting the HEALPix level.

	\begin{table}
		\centering
		\caption{The explanation of the metadata}
		\begin{tabular}{ccc}
			\hline
			
			Name & Type & Explanation \\ \hline
			RA & double & Right Ascension \\ 
			DEC & double & Declination \\ 
			DATE-OBS & timestamp & The date of the observation \\ 
			Magnitude & float & The magnitude of celestial objects \\ 
			Magnitude\_error & float & Deviation of magnitude \\ \hline
			
		\end{tabular}
	\end{table}
	
	$\bullet$ Load Balancing
	
	The partition function is intended to balance memory and workload. The problem is basically a type of N-step 0-1 knapsack problem. The 0-1 knapsack problem can be stated as
	
	\begin{equation}	
		 \hspace{3cm} max \  z = \sum_{i=1}^{n}V_{i}X_{i}
	\end{equation}
	\begin{equation}	
		\hspace{3cm}	s.t. \  \sum_{i=1}^{n}W_{i}X_{i}\leq C
	\end{equation}	
	\begin{equation}
	    \hspace{3cm} \qquad X_{i}\in \left \{ 0, 1 \right \}, 1\leq i\leq n
	\end{equation}

	where n is the number of items and the knapsack constraints with capacities C. Each item $i$ yields $V_{i}$ units of value when it occupies $W_{i}$ knapsack capacities. The goal is to find a subset of items in the backpack that yields the maximum value without exceeding the given capacities. By its nature, all entries are nonnegative~\citep{Fr2004The}.
		
	The task allocation for each process can be regarded as a solution to the 0-1 knapsack problem. The task amount that the process should be allocated to can be calculated, namely, the knapsack capacity. Each sky zone file is regarded as an item, and its volume and value are the same. Each step brings the size of the selected sky zoning files nearest to the calculated average. The capacity of the backpack is the average of the load, which is updated after each selection. The 0-1 knapsack problem can be solved by a dynamic programming algorithm. The detailed algorithm and processing flow can be found in our paper published in ISPA2017~\citep{li2017flexible}.

	\subsection{TS-Matching Algorithm}
	\label{subsec_3.3_TS-Matching Algorithm}
	
	In this study, we designed a TS-Matching calculation method for catalogues with an TS-RefTable. In addition, the overlapped indexing method is adopted to solve the boundary problems, which ensures accuracy and improves efficiency compared with the traditional method. The sky zoning file is used as input in the TS-Matching process, and the specific process is shown in Figure 4. The main features of the TS-Matching method are as follows.
	
	\begin{figure}
		\centering
		\includegraphics[width=0.45\textwidth]{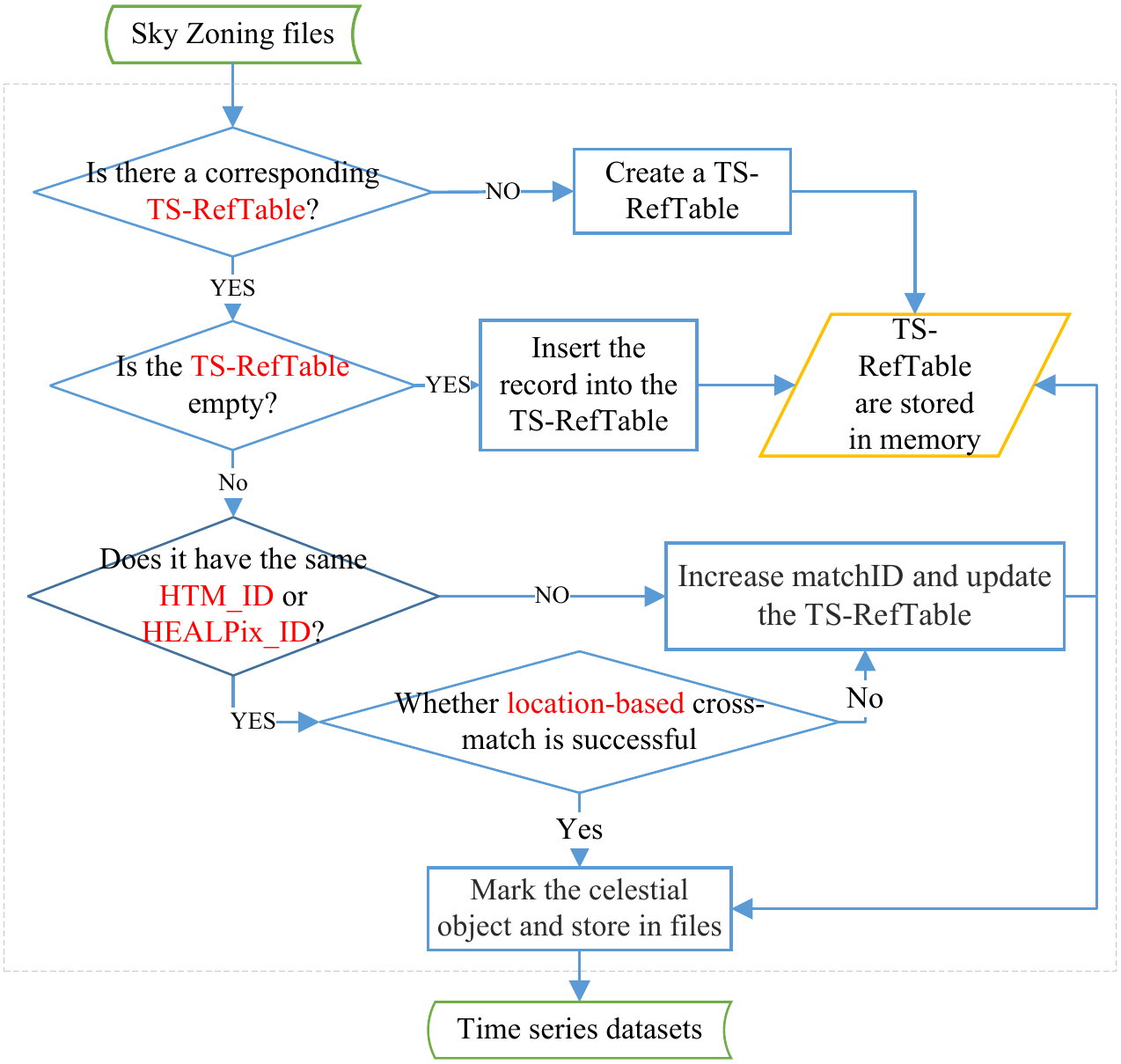}
		\caption{\small TS-Matching calculation process. Matching calculations combine TS-RefTables and location-based cross-matching. When the first record of a celestial object is produced and the TS-RefTable is empty, the celestial object is considered to be the first object and is inserted into the TS-RefTable. When the next record is produced, it is compared to the celestial object in the TS-RefTable, which has the same number in one of the indexes. If the two objects match, then they are marked as having the same match\_ID. Otherwise, the celestial object is inserted into the TS-RefTable as a new object. Finally, the marked data are stored in time series datasets.}
		\label{fig_4}
	\end{figure}
	
	$\bullet$ Overlapped Indexing Method
		
	The boundary problem is a main factor that affects the accuracy of large-scale matching calculations. Because of the errors existing in the position calibration of celestial objects, if a celestial object $A$ falls in the boundary area in a catalogue file, it is possible that its corresponding celestial object $A'$ in another catalogue file is divided into another partition. The matching calculation is only carried out between the celestial objects in the same partition. If only one index is used, the boundary problem will be serious.
		
	The common method for addressing the above boundary problems is to increase the boundary redundant data by a quick bit-operation algorithm~\citep{Zhao2009A}. However, the implementation of the traditional method is complex and inefficient. Therefore, we use the overlapped indexing method to solve boundary problems in this paper.
		
	By introducing two kinds of indexing methods, the proportion of boundary data is reduced. Because of the different shapes and levels of the two indexes, the boundary data of one index method may not be in the boundary of another index method. Only boundary data under both partitions will be lost. There are two sets of examples in Figure 5. Most of the boundary problems are solved by the method of overlapped indexing at a lower cost than the redundant method.
	
	\begin{figure}
		\centering
		\includegraphics[width=0.3\textwidth]{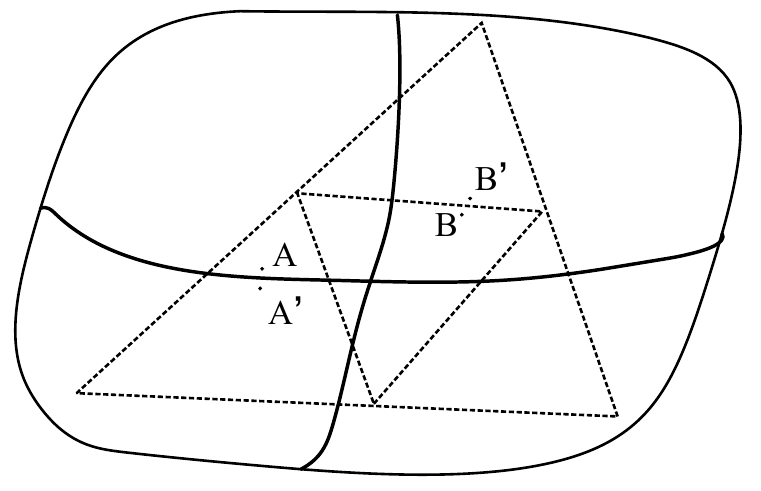}
		\caption{\small Schematic diagram of the overlapping indexing method. The solid line in the figure represents the schematic line of the HEALPix partition, and the dotted line represents the schematic line of the HTM partition. $A$ and $A'$ are two records of the same celestial object from two catalogue files, but they will be lost in the matching calculation when only using the HEALPix index without boundary redundancy processing because they are in different partitions with the HEALPix index. However, they are in the same partition as the HTM index and can be saved by the overlapped indexing method. Similarly, $B$ and $B'$ in different partitions with the HTM index, but they are in the same partition with the HEALPix index.}
		\label{fig_5}
	\end{figure}
	
	TS-Matching calculations are accelerated by using multi-processing and rely on partitioning functions in Subsection 3.2 to balance the load. The number of processes can be adjusted according to the actual situation. Partition levels are extremely important for performing TS-Matching calculations while avoiding boundary problems. High partition levels require less time, but are associated with serious boundary problems. In contrast, low partition levels can reduce the number of celestial objects, but have unacceptable response times. We address boundary problems by leveraging overlapped indexing and find the appropriate partition level according to experiments.
	
	$\bullet$ TS-RefTable
		
	We propose reference tables to effectively reduce the number of comparisons in the TS-Matching calculation. Since the read and update of the reference table are frequent, it is helpful for improving the efficiency of the whole TS-Matching process by storing it in memory rather than in databases. AstroCatR employed the in-memory reference table special designed for the TS-Matching calculation called TS-RefTable. The data structure of TS-RefTable contains positional coordinates (${RA}$, ${DEC}$), indexes (HTM\_ID and HEALPix\_ID) and match\_ID. Matching celestial objects are marked as having the same match\_ID. 
	
	The general matching calculation method needs to use relational database management system (RDBMS). According to the matching results, the data are marked and inserted into the database for later addition, deletion, modification and query. The insertion time is too long, so in the later part of the comparison experiment, it is only determined whether to store the reference table in the database as a variable for performance comparison. The design and results of the comparative experiment are shown in Subsection 4.2.
		
	The system scalability can be effectively improved by introducing an TS-RefTable. Catalogue increments can be transformed into sky zoning files. Each partition has its own TS-RefTable, which corresponds to the zone from previous TS-Matching calculations. Therefore, new incoming astronomical catalogues can be matched with the corresponding TS-RefTable to reconstruct the time series data.
	
	\section{Experiments and Results}
	\label{sec_4_Experiment and Results}
	
	To evaluate AstroCatR capabilities, we performed various experiments on real catalogue datasets during the actual 2012 AST3 observations~\citep{ma2018first}. These datasets are relatively complete observation data that are available to us currently. We adopted two types of datasets in the experiment, which are common in current time-domain astronomy. One is long interval observation of multiple sky areas (SNsurvey), the others are continuous observation of a sky area. The details of these datasets are given in Table 3. All of our experiments were executed on an Ubuntu server equipped with an Intel i7-4790 CPU (8 cores @ 3.6 GHz), 16 GB memory, and 1 TB HDD.
	
	We evaluated the ETL preprocessing of AstroCatR and how performance was affected by partition levels. Then, the performance and accuracy of reconstructing time series data were evaluated by comparing with the method using RDBMS. Eventually, we assessed the query performance of AstroCatR using executing retrieval implementations.
	
	\begin{table*}
		\centering
		\caption{The information of raw datasets in the experiment of AstroCatR}
		\begin{tabular}{lp{1.5cm}p{1.5cm}p{1.5cm}p{3cm}p{3cm}}
			\hline\noalign{\smallskip}
			Name of dataset & HD88500 & HD117688  & HD136488 & Transit & SNsurvey \\
			\noalign{\smallskip}\hline\noalign{\smallskip}
			Number of catalogue FITS files  & 591 & 655 & 660 & 3194 & 3084 \\
			Original file size (GB) & 2.5 & 6.9 & 9.0 & 101.3 & 9.6 \\
			\multirow{4}*{Statement} & \multicolumn{3}{|c|}{\multirow{4}*{HD datasets are the fields centered at these HD stars}} & A test field for detecting transit signals from exoplanets & A 500-field (\textasciitilde2000 deg$^2$) survey searching for supernovae and other transients candidates \\
			\noalign{\smallskip}\hline
		\end{tabular}
	\end{table*}
	
	\subsection{Evaluation of Preprocessing}
	\label{subsec_4.1_Evaluation of Preprocessing}
	
	We transformed catalogue FITS files from AST3 datasets into sky zoning files. We adjusted the level of indexing to control the number of sky zoning files and to prevent data from being centralized in a few files. We assessed the storage performance of the sky zoning files using the different datasets shown in Figure. 6~\citep{li2017flexible}. The size of the sky zoning files was approximately 68${\%}$ of the original catalogue FITS files regardless of the dataset. The processing times increased approximately linearly as the dataset size of the datasets grew. The processing times were acceptable for real-life usage. 
	
	\begin{figure}
		\centering
		\includegraphics[width=0.45\textwidth]{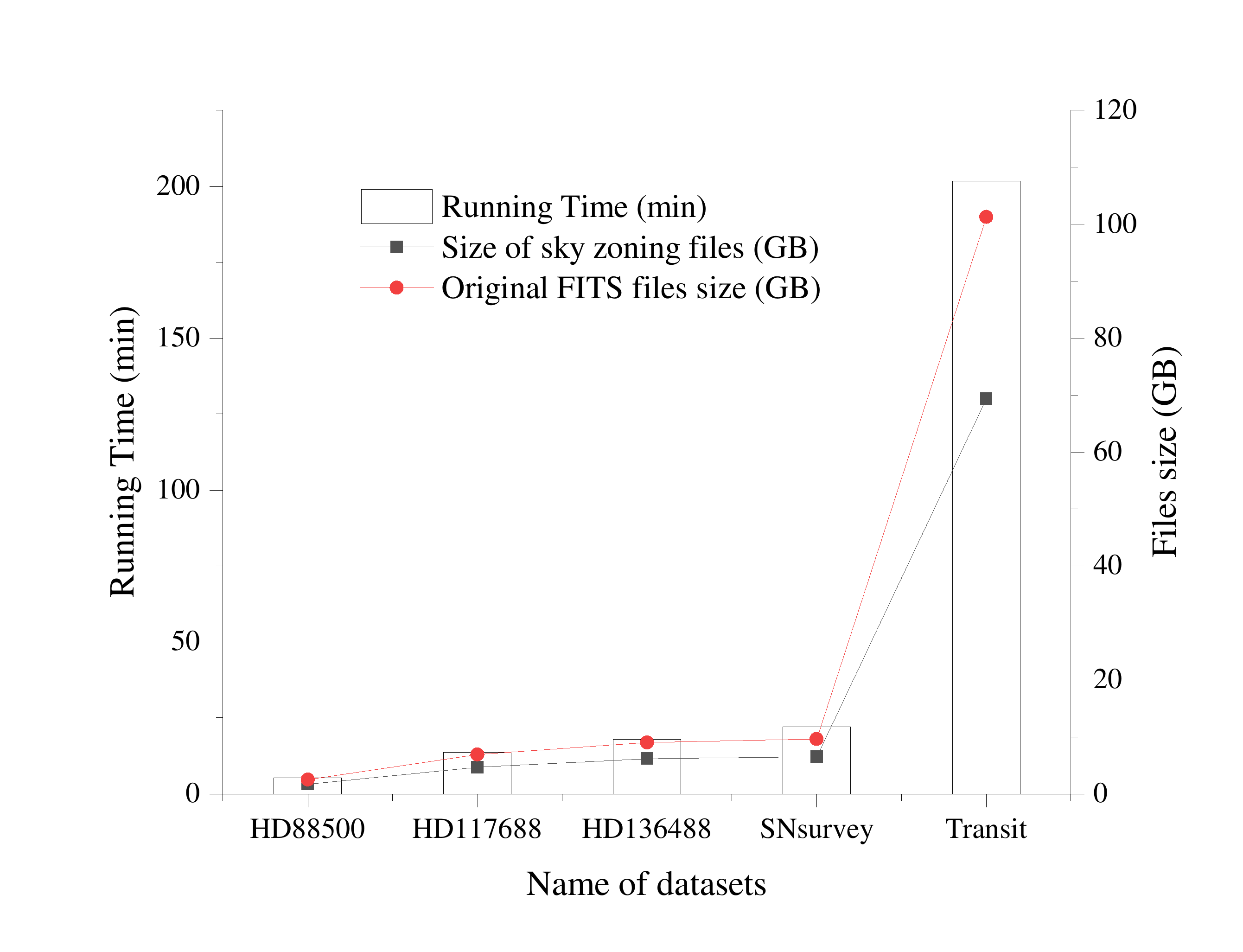}
		\caption{\small Information regarding sky zoning files. ETL preprocessing reduced storage space by nearly one-third.}
		\label{fig_6}
	\end{figure}
	
	To identify the relationship between the partition level and the number and size of the sky zoning files for ETL preprocessing, the following series experiments were conducted. The HEALPix partition level was increased from 5 to 10, so five indexes were built for each AST3 dataset. We were able to adjust the partition level according to the number of sky zoning files and the running time. We chose the AST3 datasets mentioned above for the experiment using all five indexes so that we could determine which level of partition worked best for the number of sky zoning files and running time combination. The running time and the number of sky zoning files changed with different levels as shown in Figures 7 and 8. 
	
	\begin{figure}
		\centering
		\includegraphics[width=0.45\textwidth]{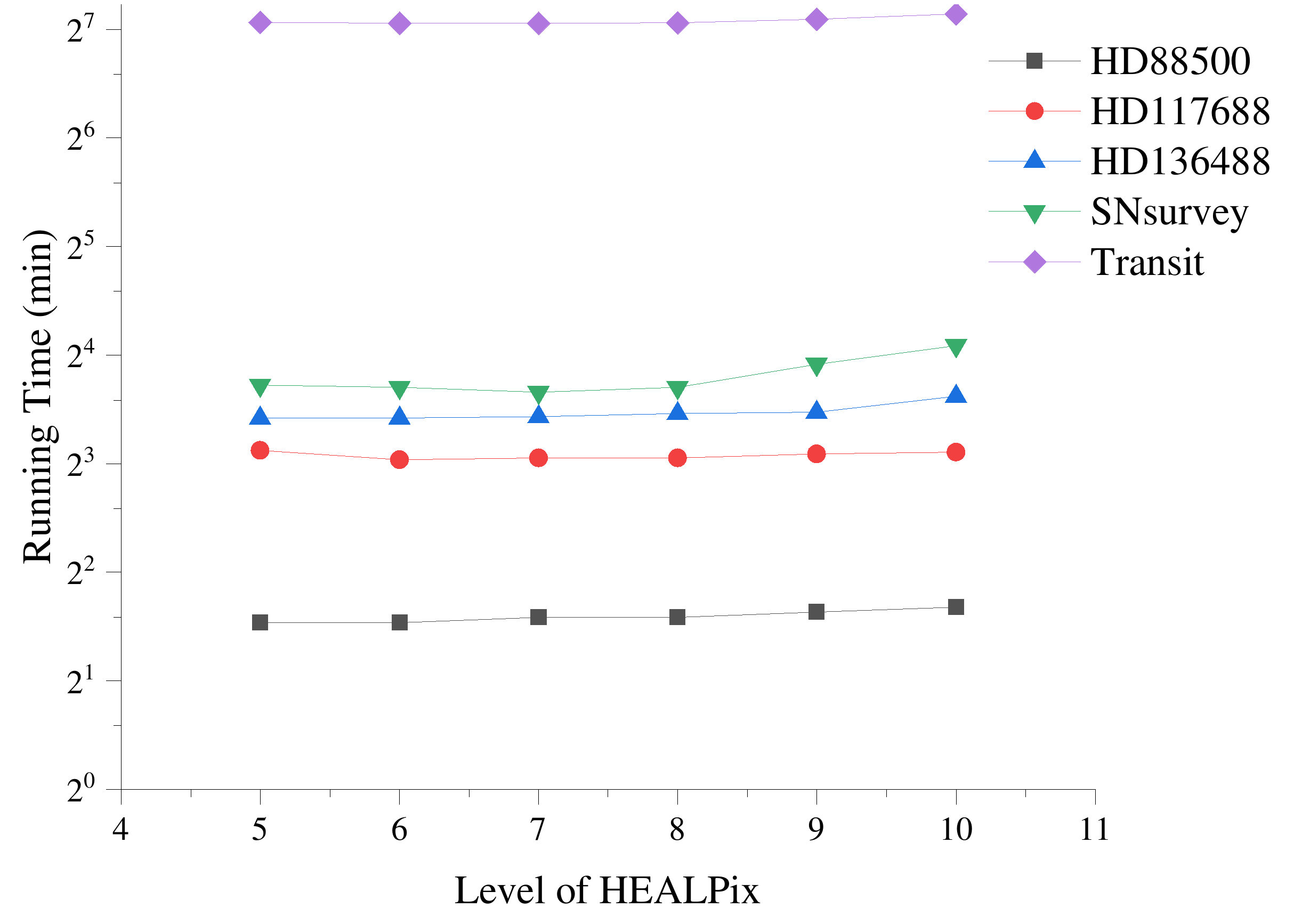}
		\caption{\small Running time (min) for different levels of HEALPix. Running time is less affected by the level of HEALPix.}
		\label{fig_7}
	\end{figure}
	
	$\bullet$ Sky Zoning File Results
	
	From the results (Figure 6), we can see that as the level increased, the running time increased because of the time required to create the files. However, the overall change was not significant. Therefore, the choice of level had little effect on ETL preprocessing performance. The number of sky zoning files increased with the level of partition, as shown in Figure 8, and more files needed to be created. The change in time consumption was not as the obvious as change in the number of files. The most suitable partition level N was in the three levels of HEALPix 6, 7, and 8. However, we needed to focus on the number of sky zoning files.
	
	\begin{figure}
		\centering
		\includegraphics[width=0.45\textwidth]{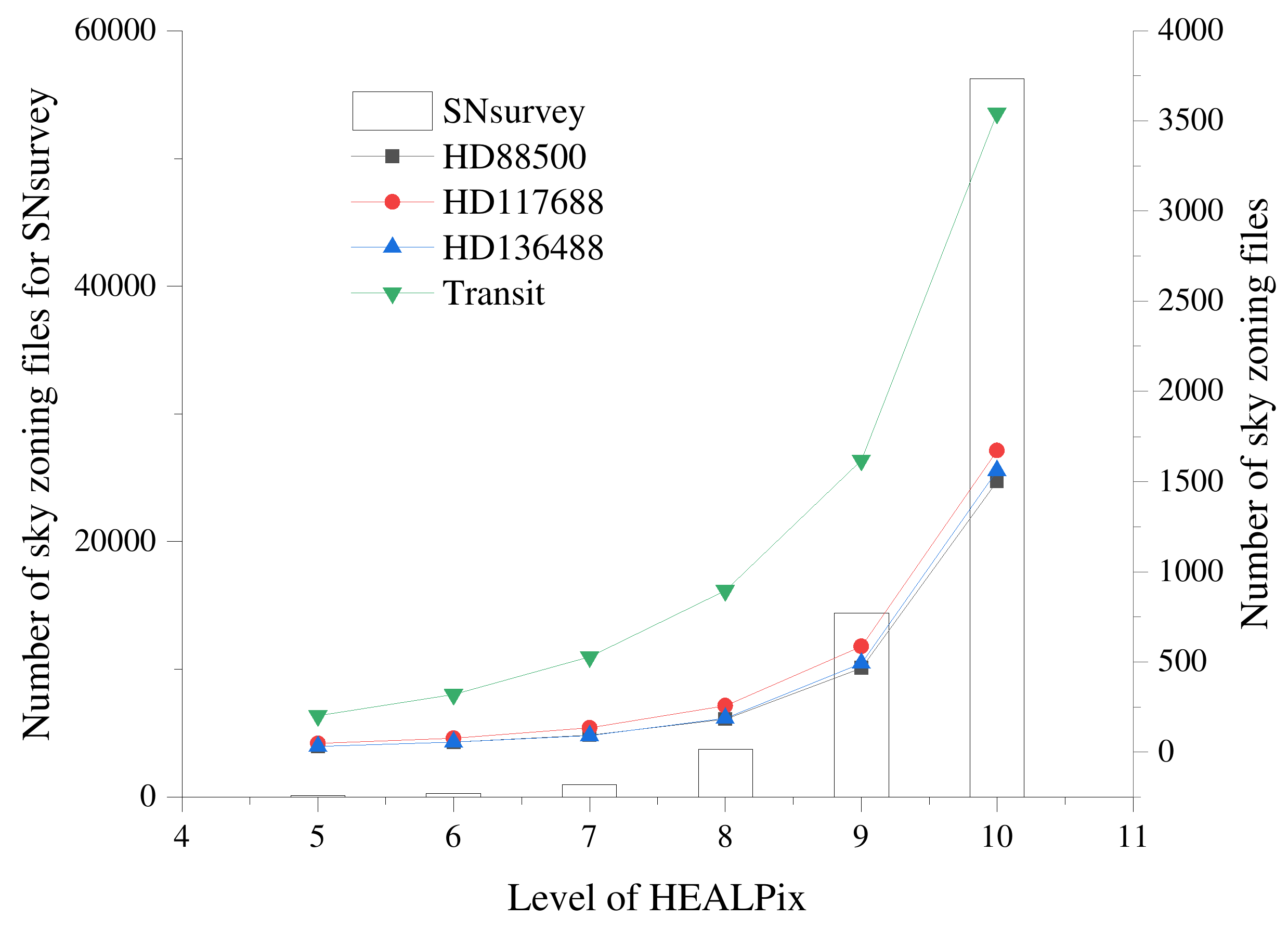}
		\caption{\small Number of sky zoning files for different levels of HEALPix. The level of HEALPix had approximate effects on catalogue data.}
		\label{fig_8}
	\end{figure}
	
	$\bullet$ Partition Function Performance
	
	To ensure appropriate load balancing after partitioning is performed, it is necessary to avoid centralizing large percentages of the data into a few sky zoning files. Therefore, the number of sky zoning files should be as moderate as possible. Figure 8 shows that the number of files increases with the level. The sky zoning files generated by the SNsurvey dataset are too large, which seriously affects the performance observations of other datasets at different partition levels. As a result, the SNsurvey dataset is assigned a separate axis. We needed to select a level to produce a moderate number of sky zoning files, and distribution of data that was as even as possible. According to Figure 7, the most appropriate partition level (N) for the four datasets was 6 or 7.
	
	\subsection{Reconstructing Time Series Performance}
	\label{subsec_4.2_Reconstructing Time Series Performance}
	
	Prior to this study, the general approach was to insert the original astronomical catalogues into RDBMS and mark celestial objects according to the results of matching calculations. MySQL and PostgreSQL are two representative relational database management systems. All of the experiments in this subsection were executed on an Ubuntu server equipped with an Intel i7-4710 CPU (8 cores @ 2.5 GHz), 8 GB memory, and 500 GB HDD.
	
	The general method for reconstructing time series data is to use RDBMS. The matching procedure resembles the approach described earlier. The reference table is stored by the table of RDBMS, and it consists of ${RA}$, ${DEC}$, HTM\_ID, HEALPix\_ID, and match\_ID. The same celestial objects have the same match\_ID. The HTM\_ID and HEALPix\_ID in the table are indexed. The record of celestial objects is marked by the result of the matching calculation and inserted into the data files. Even if batch insertion technology accelerates the data insertion speed, its time consumption is very large. Therefore, the processed data were not inserted into the databases, and we do not discuss the retrieval of time series data in the databases.
	
	$\bullet$ Accuracy
	
	To verify the accuracy of the results, we compared the experimental results of three groups of experiments. The main goal was to ensure that the produced matching was correct. Therefore, we counted the numbers of celestial objects in the reference tables of these experiments. They were the same when processing the same dataset. Furthermore, we conducted manual verification experiments with small samples. The results of cross-matching based on location were the same as those produced by AstroCatR.
	
	$\bullet$ TS-RefTable Versus RDBMS-based reference table
	
	To evaluate the performance of reconstructing time series data, we compared it with the method using RDBMS, as shown in Figure 9. The processing time includes the TS-Matching calculation of sky zoning files. The method using RDBMS required too much time for large columns of datasets. Therefore, the experiments were performed using a set of catalogue FITS files. The experimental results of AstroCatR were single-process. The method using RDBMS required significantly more time than the AstroCatR method. There were two types of relational databases selected for use, MySQL and PostgreSQL. The reference table was stored in the memory table of MySQL. Although PostgreSQL was used for unlogged tables, its insertion speed was slower than MySQL. The AstroCatR method was demonstrably more efficient; according to all six experiments, it ran in only approximately 30${\%}$ of the time of the methods using RDBMS. 
	
	\begin{figure}
		\centering
		\includegraphics[width=0.45\textwidth]{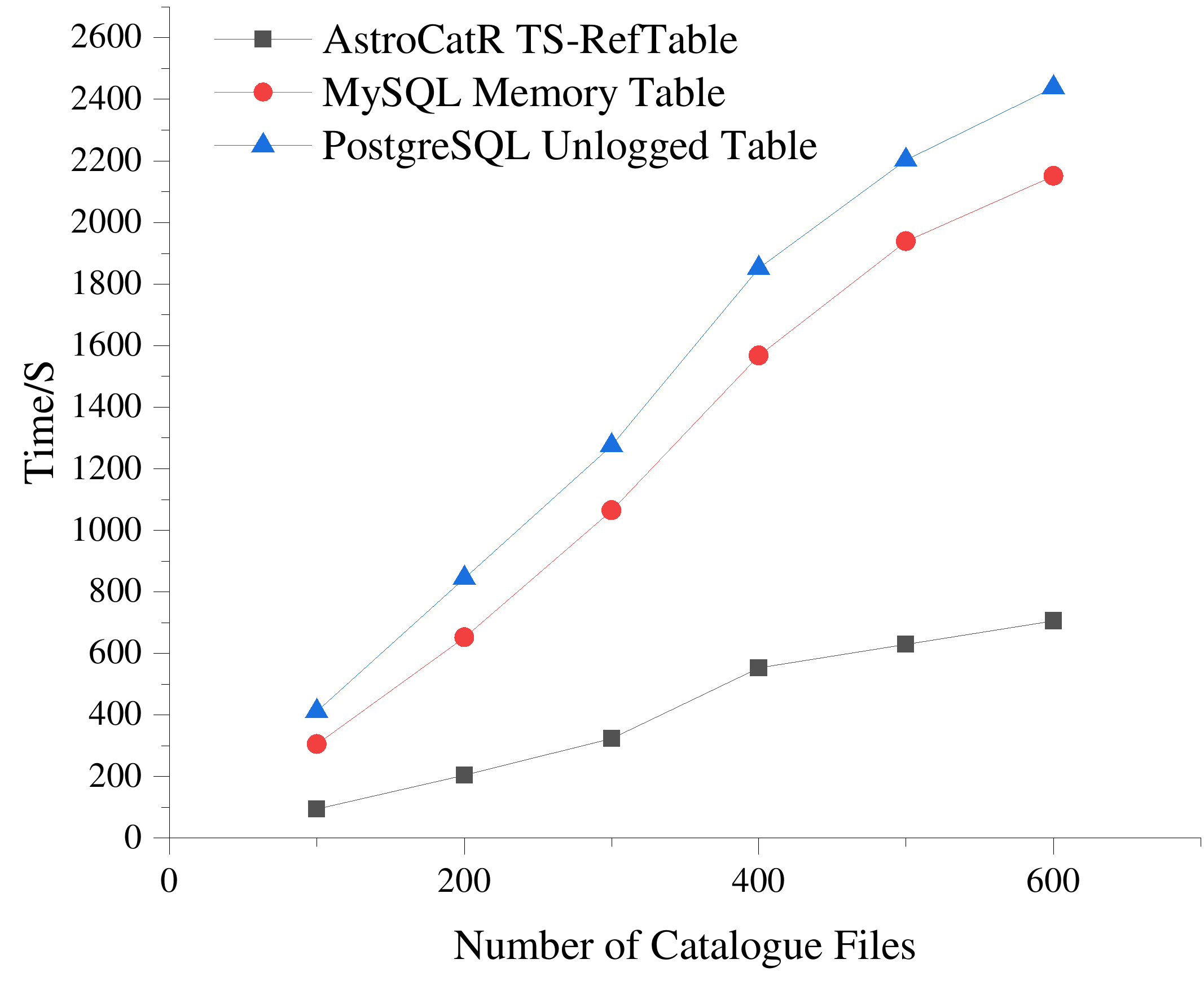}
		\caption{\small Performance of the AstroCatR TS-RefTable, MySQL memory table, and PostgreSQL unlogged table. The TS-Matching calculation of AstroCatR is nearly 3X faster than RDBMS.}
		\label{fig_9}
	\end{figure}
	
	\begin{figure}
		\centering
		\includegraphics[width=0.45\textwidth]{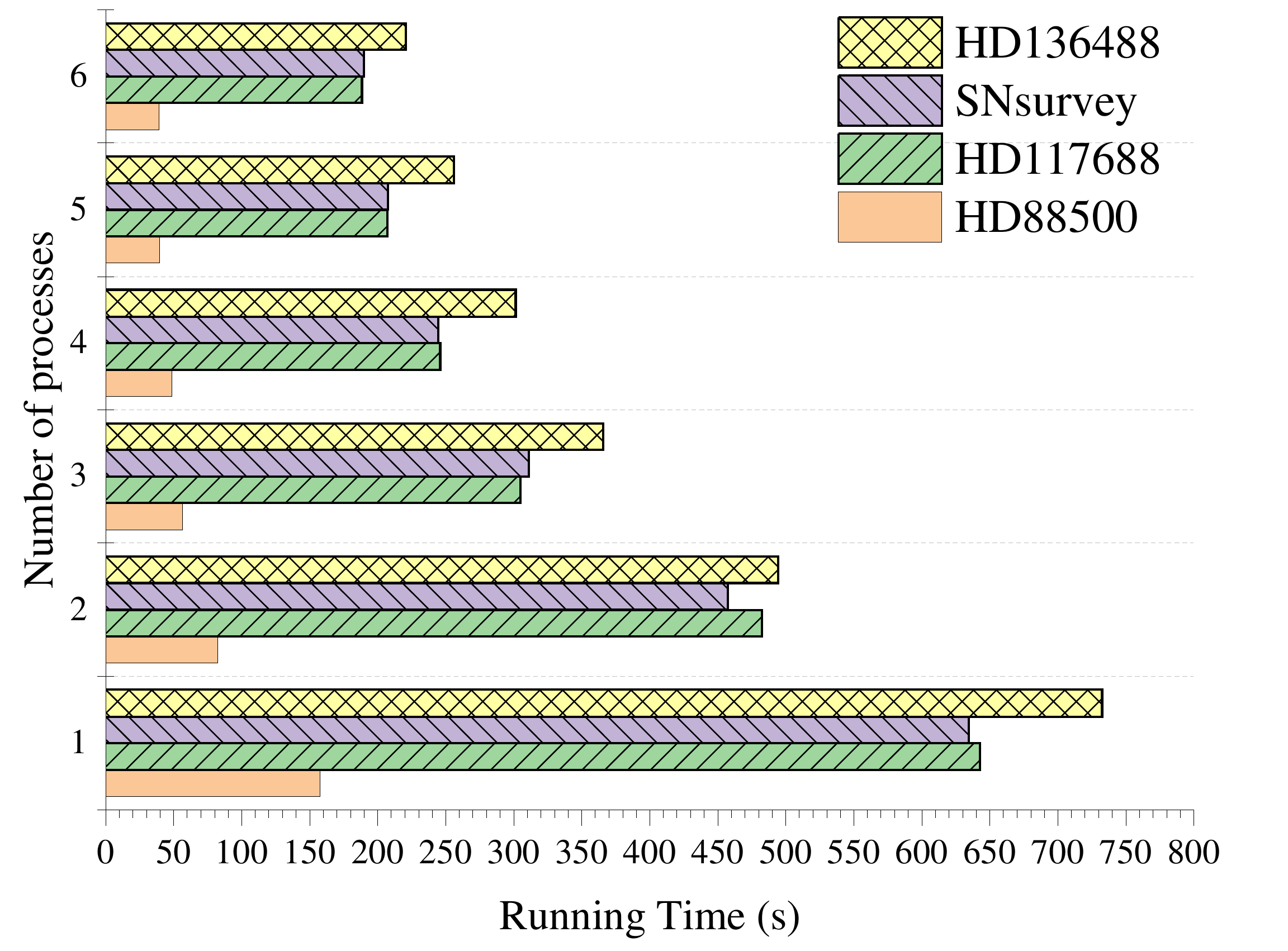}
		\caption{\small Performance of AstroCatR for different numbers of processes. As the number of processes increased, the effects of acceleration began to appear.}
		\label{fig_10}
	\end{figure} 
	
	$\bullet$ Parallel Processing Performance
	
	To analyze the performance of AstroCatR for different numbers of processes, we chose several datasets on which to perform TS-Matching experiments, and plotted the results in Figure 10. There is a big gap between the data volume of the transit dataset and other datasets. Therefore, the results of datasets with similar size are selected for display. From the results, we can see that the running time decreased as the number of processes increased. Because of the performance of the partition function, the workload of each process was balanced, and the time consumption of the partition function was negligible relative to the time required for the subsequent TS-Matching calculations~\citep{li2017flexible}. However, this reduction was limited by the communication between processes. The running time increased with the size of the dataset for the same number of processes. However, the time was reduced for the SNsurvey dataset because the number of comparisons was fewer than it was for other datasets. The speedup ratio increased steadily as the data volume increased.

	\subsection{Querying Performance}
	\label{subsec_4.3_Querying Performance}
	
	In this set of experiments, we evaluated the retrieval of time series. Because of the processing partition, the actual query was completed in two steps: data location and data query. First, we calculated the number of sky zones corresponding to ${RA}$ and ${DEC}$ and determined the specific partition of the data. Then, all of the time series data for the partition and corresponding light curves were provided to the user. 
	
	Figure 11 shows the query response time of AstroCatR for different numbers of celestial objects. We found that the time increased with increases in the number of celestial objects. The time overhead for the query mainly depends on the size of the located sky zoning file because the time series data of the target objects must be extracted from that file.
	
	\begin{figure}
		\centering
		\includegraphics[width=0.45\textwidth]{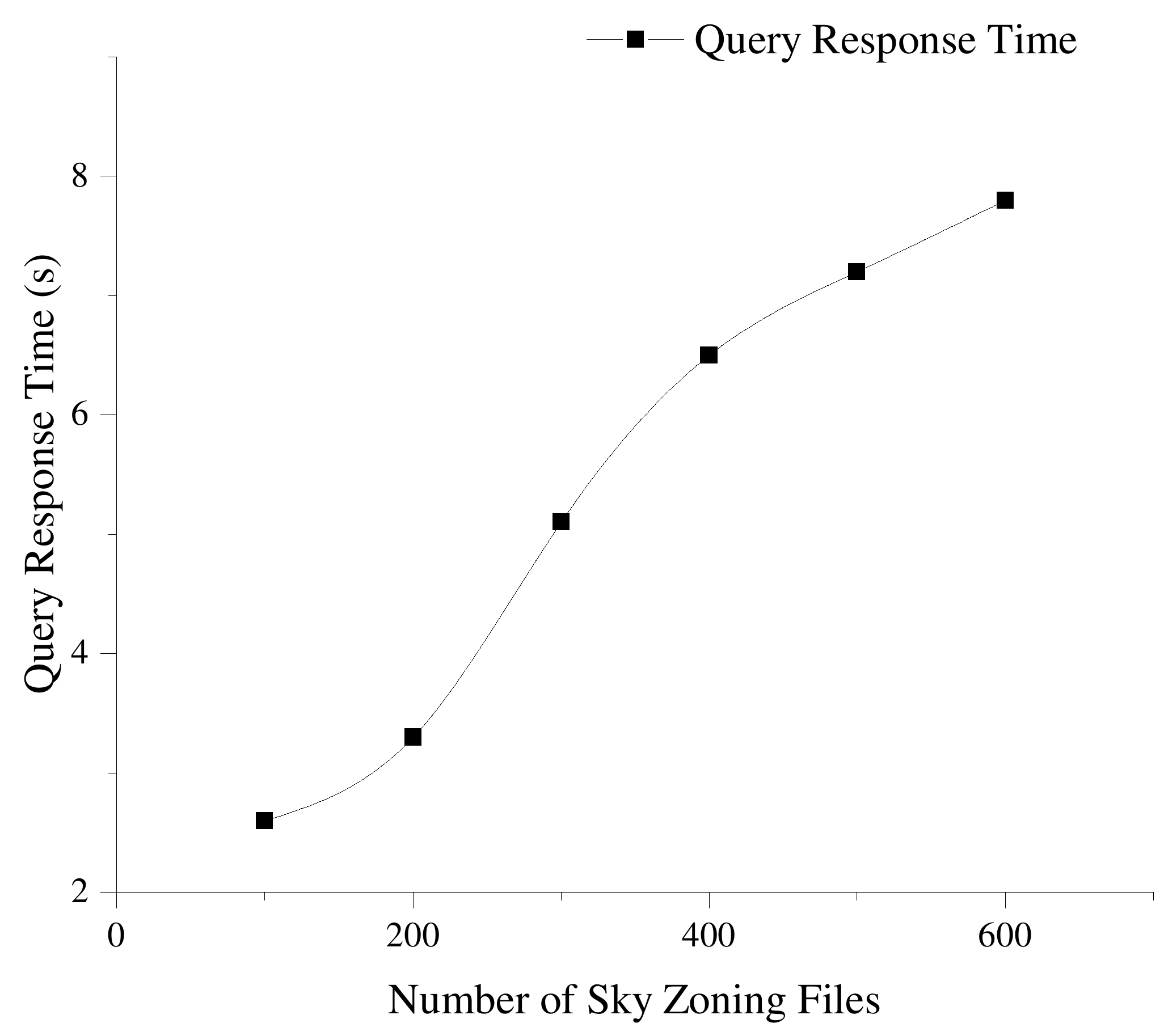}
		\caption{\small Query performance. Response times increased with the number of files. Time consumption was primarily affected by the size of the sky zoning file being located.}
		\label{fig_11}
	\end{figure}
	
	\begin{figure*}
		\centering
		\includegraphics[width=0.95\textwidth]{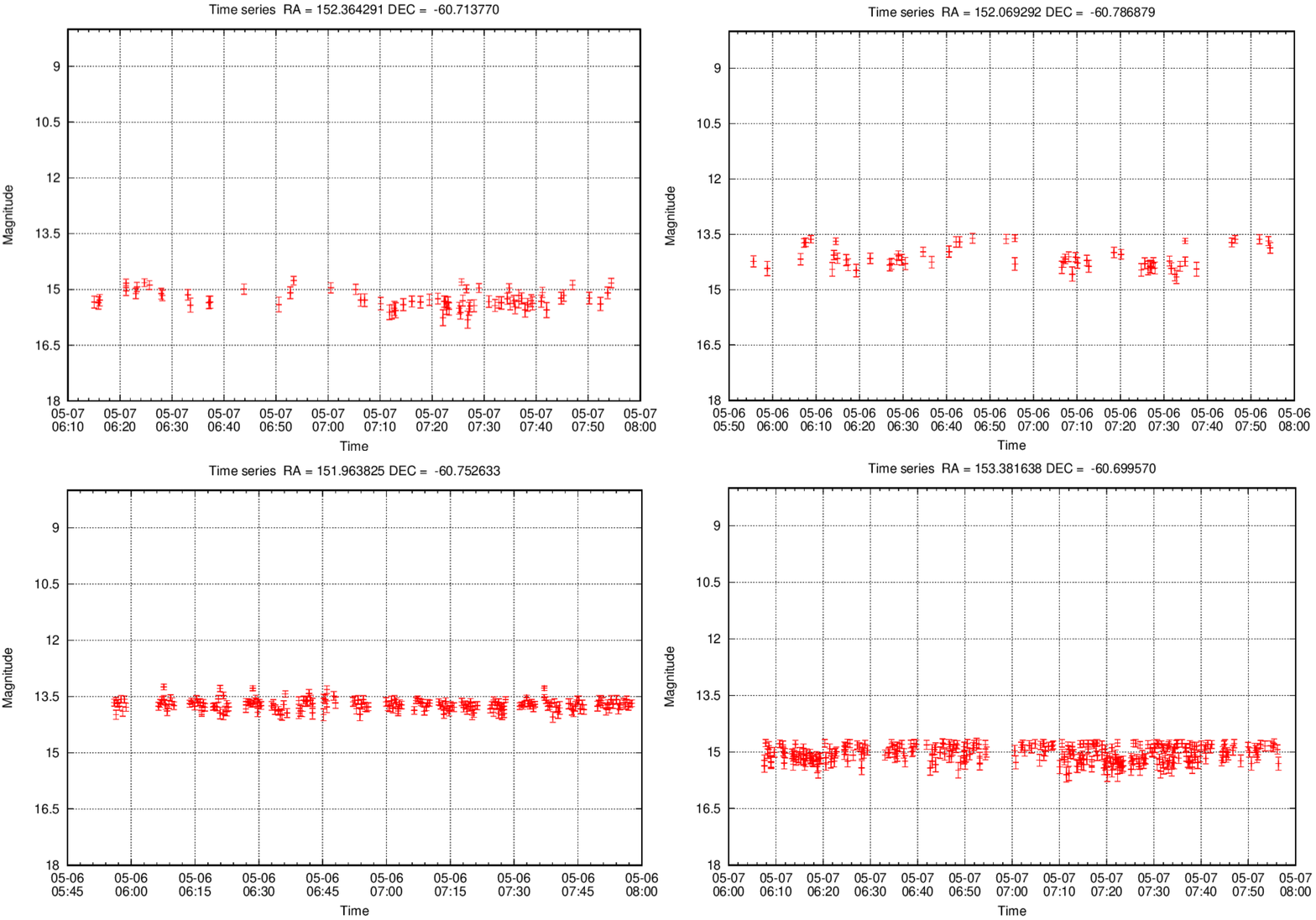}
		\caption{\small Samples of light curve (scatter plots).}
		\label{fig_12}
	\end{figure*}
	
	Figure 12 illustrates several light curves created by Gnuplot according to the time series data in the obtained sky zoning file. The X-axis represents the observation time, and the Y-axis represents the magnitude. In addition, the vertical red lines indicate the error estimates for the magnitude. Time series of celestial objects and their corresponding data support the study of TDA in areas such as time series prediction, exoplanet transit classification, and detection.

	\section{Conclusion and Future Work}
	\label{sec_5_Conclusion and Future Work}
	In this study, we proposed AstroCatR, an efficient and scalable system for reconstructing time series data from astronomical catalogues. Several datasets provide light curves for celestial objects, such as Gaia. AstroCatR is intended to provide researchers with the opportunity to analyze the time series data of each catalogued celestial object to discover valuable missing information. Users can choose a position coordinate to obtain the time series data and corresponding light curves. Additionally, AstroCatR provides flexibility for users to increase special information associated with data by modifying configuration files.  
	
	Furthermore, users can set the number of processes and levels of partitions according to the actual situation and achieve the desired performance based on the partition function. We leveraged a novel approach to accelerate and promote the accuracy of TS-Matching calculations using TS-RefTable and overlapped indexing. AstroCatR can efficiently and flexibly process data increments and perform TS-Matching calculations on TS-RefTables. The experimental results of the TS-RefTable and traditional RDBMS-based reference table show that AstroCatR is three times faster in processing large-scale astronomical catalogues.
	
	In future research, we will work on optimizing time and boundary problems. We provide time series data for candidate celestial objects, whose accuracy needs further research to prove, but we will ensure the accuracy of these candidates as much as possible. We can make a new attempt at indexing, dividing the sky area, and dealing with boundary problems, reducing the number of TS-Matching calculations, and facilitating the reconstruction process of the time series data from catalogues. We will try to analyze the time series data through AI technology for obtaining as many candidates as possible with scientific goals. 
	
	\section*{Acknowledgments}
	 This work is supported by the National Natural Science Foundation of China (11803022), the Joint Research Fund in Astronomy (U1731243, U1931130, U1731125) under cooperative agreement between the National Natural Science Foundation of China (NSFC) and Chinese Academy of Sciences (CAS). Data resources are supported by China National Astronomical Data Center (NADC) and Chinese Virtual Observatory (China-VO).
	%Here are two sample references: \cite{Feynman1963118,Dirac1953888}.
	
	\section*{Conflicts of interest}
	The authors declare no conflict of interest.

	
	
	\bibliographystyle{mnras}
	\bibliography{AstroCatR} % if your bibtex file is called example.bib

	%%%%%%%%%%%%%%%%%%%%%%%%%%%%%%%%%%%%%%%%%%%%%%%%%%
	
	%%%%%%%%%%%%%%%%% APPENDICES %%%%%%%%%%%%%%%%%%%%%
	
	%	\appendix
	%	
	%	\section{Some extra material}
	%	
	%	If you want to present additional material which would interrupt the flow of the main paper,
	%	it can be placed in an Appendix which appears after the list of references.
	
	%%%%%%%%%%%%%%%%%%%%%%%%%%%%%%%%%%%%%%%%%%%%%%%%%%

	% Don't change these lines
	\bsp	% typesetting comment
	\label{lastpage}
\end{document}